\documentclass[doublecol]{epl2} 
\usepackage{amssymb, amsmath}
\usepackage{bm}
\usepackage{graphicx}
\usepackage{color}
\usepackage{amsfonts}
\usepackage{latexsym}
\usepackage{enumerate}

\newcommand{\hh }[1]{ \hat{\bm{#1}} }
\newcommand{\m }[1]{ \mathbf{#1} }

\title{ {     Dynamics and  interactions of active rotors} }

\shorttitle{Dynamics of active rotors} 

\author{M. Leoni \and T. B. Liverpool}

\institute{Department of Mathematics, University of Bristol, Clifton, Bristol
BS8 1TW, U.K.}
\pacs{47.63.mf}{Low-Reynolds-number motions}
\pacs{05.65.+b}{Self-organized systems}
\pacs{87.10.-e}{General theory and mathematical aspects}

\abstract{     
We consider a simple model of an internally driven  self-rotating object; a rotor,
 confined to two dimensions by a thin  film of low Reynolds number fluid. 
We undertake a detailed study of the hydrodynamic interactions between a pair of rotors and find that their effect on the resulting dynamics is a combination of fast and slow motions.  We analyse the slow dynamics using an averaging procedure to take account of the fast degrees of freedom. Analytical results are compared with numerical simulations. 
Hydrodynamic interactions mean that while isolated rotors do not translate,  bringing together a pair of rotors leads to motion of their centres. 
Two rotors spinning in the same sense rotate with an approximately constant angular velocity around each other, while two rotors of opposite sense, both translate with the same constant velocity, which depends on the separation of the pair.  As a result a pair of counter-rotating rotors are a promising model for controlled self-propulsion in two dimensions. 
 }

\begin{document}

\maketitle

Soft active systems are composed of interacting units that consume energy and generate local motion and mechanical stresses.   They show a rich variety of collective behaviour, including dynamical order-disorder
transitions and pattern formation on various scales.  
A much studied example are collections of self-propelled objects
~\cite{ Pedley,  Wu,
Rama, TYR05, Domborowski, Dreyfus, Sokolov, Saintillan, Peruani, Chate, Leptos, Gosh, Aparna,kruse,ahmadi} which independently translate due to internally generated motions by the objects themselves.

In this article we study, a particularly simple class of self-driven particles and their interactions. These are self-rotating objects (that we call rotors) lying in a plane~\cite{ LJJP, Pago,Uchida}.  
Our motivation is two-fold. 
First recent developments~\cite{CSS, dhar, riedel,grzyb} have shown the possibility of
experimentally realising  such self-driven rotors on the nano-scale.  
In addition recent progress in nano-scale microfabrication techniques 
have been able to generate synthetic chemically driven 
rotors that rotate fast enough so that the effects of their interactions on their dynamics 
can been measured experimentally~\cite{2rotors}. 
Second, a collection of active rotors provide a particularly simple example of an active system as their collective behaviour can be described in terms of 
scalar rather than vector equations. 
The first step in the quest to understand their collective behaviour is to develop a detailed understanding of their interactions. 

Motivated by this we investigate analytically and numerically  the 
dynamics of rotors 
confined in two
dimensions (the $x-y$ plane) by a viscous film, such as a membrane.
{We  first introduce the model of a single rotor: (which when isolated rotates around its centre but does not translate).  Each rotor can be described by  a scalar $\sigma$, the value of the projection of its  angular velocity $\bm{\omega}$ on the $\hh z$ direction, that we call \emph{spin}. We  next  study the hydrodynamic interactions between a pair, in the absence of noise when the distance between their centres is large compared to their size.  We 
 find that the resulting dynamics  can be described in terms of slow and fast variables that depend on the relative spin of the pair.  We obtain analytic expressions for the dynamics of the slowly varying quantities and find that}
 for opposite spins the hydrodynamic interactions lead to a net translation of the pair while for like spins they lead to a relative rotation of the pair around each other.  
{     These leading order analytic results are in good agreement with  numerical simulations for well separated rotors. 
Interestingly, our study suggests that a pair of counter-rotating rotors could be used to construct  a 
 micron sized-swimmer in two dimensions, with the speed of the swimmer controlled in a precise manner by varying the equal and opposite torques acting on the rotors.
 }

\section{Hydrodynamics of a thin fluid film}
\begin{figure}[!ht]
\centering
\includegraphics[width=0.25\textwidth]{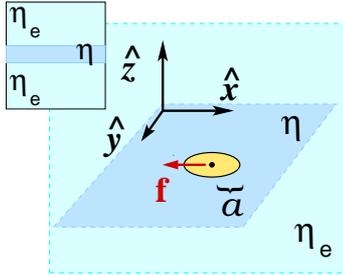}
\caption{ Thin film. A two dimensional fluid layer characterized by a
2d viscosity $\eta$ fills the whole plane $z =0$. It is surrounded, above and
below, by another fluid having a 3d viscosity $\eta_e$ with which it is
coupled hydrodynamically. Objects are constructed out of groups of 
disks of radius $a$ that are subject to forces lying entirely in the plane $
z =0$.}  \label{fig:membrane}
\end{figure}

The active rotors can be confined to two dimensions by being adsorbed onto a thin fluid film
{  
~\cite{SD75, Huges, Goldstein, NLP07, LJJP, LLM}.  
}
The film can be described as an infinite incompressible two dimensional layer of  fluid with (2d)  viscosity
$\eta$ filling the plane $z =0$ 
and coupled hydrodynamically to another incompressible bulk  fluid of  (3d) viscosity $\eta_e$ which
fills the region $z \neq 0$ (see Fig.~\ref{fig:membrane}).
We consider  the fluid dynamics in the  vanishing Reynolds number (Stokes) limit where inertia can be neglected~\cite{Lifshitz:1987jw}. Given a force density ${\bm F}(x,y)$, the in plane equation reads 
\begin{equation}
\eta \nabla_\perp^2 \bm{v} + \nabla_\perp p +  \sigma^+_e-\sigma^-_e = -\bm{F} \; ; \; \nabla_\perp \cdot \bm{v}=0
\label{eq:in-plane}
\end{equation}
where $\bm{v}(x,y)$ is the 2-dimensional velocity field (in the plane $z=0$), $\nabla_\perp=(\partial_x,\partial_y)$  is the 2-d gradient operator, $p(x,y)$ is the in-plane pressure. $\sigma^{\pm}_e := \eta_e  \partial_z\bm{v}'\vert_{0^{\pm}} $ is the shear stress of the bulk fluid at the top/bottom of the thin film; $\bm{v}'(x,y,z)$ is the 3-dimensional velocity field defined in the external region $z\neq 0$ which satisfies the Stokes equation 
\begin{equation} 
\eta_e \nabla^2 \bm{v}' + \nabla p' = 0 \quad ; \quad \nabla \cdot \bm{v}'=0
\end{equation}
where $\nabla=(\nabla_\perp,\partial_z)$  is a 3-dimensional gradient operator  and $p'(x,y,z)$ is the bulk pressure.
The ratio of the two and three dimensional viscosities introduces a length-scale $l := \eta/2
\eta_e$.  Associated with the existence of this length-scale  are two
asymptotic regimes: at lengths $r \gg l$ dissipation is mostly due to flow out
of plane, and the hydrodynamics is similar (but {\em not} identical) to that in three dimensions;
while for $r \ll l$ dissipation occurs almost entirely in plane and thus
hydrodynamic flow fields are quasi-two dimensional.
We construct objects out of  collections of flat disks  of radius $a$ 
subject to point forces $\mathbf{f}$ at their centres lying entirely in-plane. 
The viscous drag on a disk moving at a constant speed can be obtained in an
analogous way to the Stokes' solution for a sphere in three dimensions and the
resulting drag has the form
{  
~\cite{SD75, Huges, NLP07}
}
\begin{math}
 \gamma = 4 \pi \eta/g
\label{eq:gamma}
\end{math}
where $g$ is a function of $l/a$.

The interaction between disks embedded in the film can be obtained using the Green function $\bm{H}$ of eq.~(\ref{eq:in-plane}), corresponding to the flow, $ \bm{v}(\m r) = \bm{H}(\m r - \m r_0) \cdot {\bf f}_0$ generated by an {\em in-plane}  point-like force, or stokeslet, ${\bf f}_0$ at ${\bf r}_0$.
This description is well suited in the limit where the separation between  disks is much greater than their radii so they can be approximated as point-like objects. The tensor $\bm{H}({\bf r})={l \over \eta}\int {d^2 k \over (2 \pi)^2} e^{-i {\bf k \cdot r}} {\left({\Bbb I} - \hat{\bm k} \otimes \hat{\bm k} \right) \over k^2 + k l}$ is the thin film equivalent of the Oseen tensor
{ 
~\cite{Doi,LLM} 
}
and we refer to it as the hydrodynamic tensor.
In the limit $r \gg l$,
\begin{equation}
\bm{H}(\mathbf{r}) = \dfrac{l}{2 \pi \eta}  \dfrac{\hh r \otimes
\hh r}{r} + O\left(({l / r})^2\right) .
\end{equation}
For a set of $N$ discs with forces ${\bf f}_k$ at their centres, ${\bf r}_k$ ($k \in \{1,N\}$) each disk is subject to
 the dynamic equation
\begin{equation}
 \m v_n =   \sum^{N}_{m, n = 1 } \bm{\mathcal{H}}_{nm} \cdot \m f_m
\end{equation}
 where, for $n \neq m$ $\bm{\mathcal{H}}_{nm} := \bm{H}(\m r_n -\m r_m)$, and for $n = m$ we have $\bm{\mathcal{H}}_{nn} := \mathbb{I}/\gamma$. 
 This allows us to characterise the flow fields on scales much larger than $a$. 
 
\section{The rotor model }
\begin{figure}[!ht]
\centering
\includegraphics[width=0.25\textwidth]{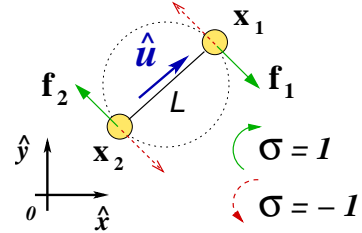}
\caption{Instantaneous configuration of a single rotor. For fixed $f$, there are 2 possible
choices for the force distribution, determined by $\sigma$,  as illustrated,
resulting in two different senses of rotation. 
We associate $\sigma = 1$ with clockwise rotation and vice-versa for $\sigma =
-1$.}  \label{fig:rot-swimm-cg-ntf}
\end{figure}

A rotor is composed of  two disks, labelled  with the index $n=1,2$. 
The disks are a fixed distance $L$ apart, oriented along the direction $\hh u$ (lying in-plane). Denoting by $\m x_n$ the position of the disk $n$,
{ $L \hh u =\m x_1-\m x_2$}.
Equal and opposite point forces  (also lying in-plane) act at the centre of the disks  directed perpendicular to their separation, applying a torque 
on the rotor. {The associated  force density  is 
 $ \m f_n\delta(\m x -\m x_n)$, where $\m f_n := (-)^{n+1}\sigma f \hh u^{\perp} $
so that $ \m f := \m f_1 = - \m f_2  $. }
The magnitude of the rotor torque, $fL$ and the spin
$\sigma=\pm 1$ (the sense of rotation), parametrise the rotor.
The director $\hh u^{\perp}$ is obtained from $\hh u$ by a clockwise rotation of an angle
$\pi/2$. See Fig.~\ref{fig:rot-swimm-cg-ntf} for an illustration.
{Neglecting thermal noise,}
 the equations of motion are
\begin{math}
 \dot{\mathbf{x}}_1 = \left( {\Bbb I} /\gamma - 
\bm{\mathcal{H}}_{12} \right) \cdot \mathbf{f} \; , \; 
 \dot{\mathbf{x}}_2 = - \left(  {\Bbb I}/\gamma - 
\bm{\mathcal{H}}_{21} \right) \cdot \mathbf{f} \; ,
\end{math}
{where $\bm{\mathcal{H}}_{12} = \bm{\mathcal{H}}_{21} = \frac{l}{2 \pi \eta} \frac{\hh u \otimes \hh u}{L}$.}

By construction, the rotor centre
 {  $\m R =\frac{1}{2} (\m x_1 + \m x_2)$}  
is stationary while 
 {the orientation axis  is not.  Its equation $\dot{\hh u} = \frac{2 \m f}{L \gamma}$, describes  rotation  with constant angular velocity around the centre.
To see this it is convenient to  introduce  the normal, $\hh z$ defined as $\hh u \wedge \hh u_\perp = \hh z$,  a unit vector in directed along the negative $z-$axis. 
The equation for $\hh u$ can then be recast as: 
\begin{equation}
\dot{\hh u} = \bm{\omega}_0  \wedge \hh u 
\label{eq:omega0}
\end{equation}
 with angular velocity, $\bm{\omega}_0 :=   \frac{2f}{L \gamma} \sigma \hh z$.  
 The period of the rotor  is given by $T_0 := 2 \pi / \omega_0$, where $\omega_0 = 2 f/(L \gamma)$.
The spin $\sigma$ is simply the projection of the angular velocity director $\hh
\omega_0$ on to  $\hh z$. 
Equation~(\ref{eq:omega0}) determines $\hh u(t)$ up to a phase. We choose $\hh u = \hh x \sin(\sigma \omega_0 t) + \hh y \cos(\sigma \omega_0 t) $ and $\hh u^\perp = \hh x \cos(\sigma \omega_0 t) - \hh y \sin(\sigma \omega_0 t) $.
 }

\section{Dynamics of two interacting rotors}
\begin{figure}[!ht]
\centering
\includegraphics[width=0.3\textwidth]{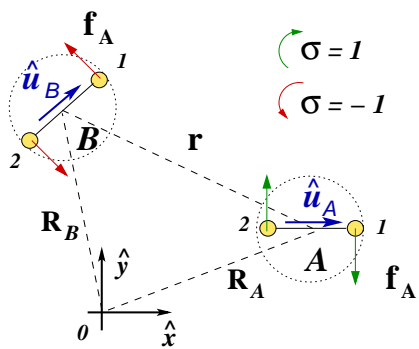}
\caption{ Instantaneous configurations of two interacting rotors. The two
rotors, labelled with the letters $A,B$, lie in the plane at a distance $r$
between their centres. The spin values are $\sigma_A =1$ and $\sigma_B =-1$.}  \label{fig:2rot-cg-ntf}
\end{figure}

It is straightforward to generalize the approach above to deal with two rotors, which we label $A$ and $B$  (see Fig. \ref{fig:2rot-cg-ntf}).
There are now 4 disks with centres at positions, $\m x^{\alpha}_n$, where $n =1,2$ and $\alpha = A, B$. 
It is convenient to describe the dynamics in terms of orientations $\hh u_{\alpha}$ and the centre positions $ \mathbf{R}_{\alpha}$ of each rotor, defined by 
\begin{math}
L \hh u_{\alpha}  := \mathbf{x}^{\alpha}_1 - \mathbf{x}^{\alpha}_2 \;  ;  \; 
\mathbf{R}_{\alpha}  := \dfrac{1}{2}( \mathbf{x}^{\alpha}_1 +
\mathbf{x}^{\alpha}_2) \; , 
\end{math}
respectively.
A single rotor cannot propel itself through the fluid. 
Interestingly, however,  net motion of the rotor center  occurs
when another rotor is close by. This can be seen in the equations of motion for $\m R_{\alpha}$ and $\hh u_{\alpha}$
{that, in absence of noise, are} :
\begin{align}
& \dot{\hh u}_{\alpha}   =  \bm{\omega}^\alpha_0 \wedge \hh u_{\alpha}  + \big(
\bm{\mathcal{H}}_{1 \alpha 1 \beta} - \bm{\mathcal{H}}_{1 \alpha 2 \beta} -
\bm{\mathcal{H}}_{2 \alpha 1 \beta} + \bm{\mathcal{H}}_{2 \alpha 2 \beta} \big) \cdot 
\dfrac{\m f_{\beta} }{L};  \nonumber \\
\nonumber \\
& \dot{\mathbf{R}}_{\alpha}  = \dfrac{1}{2}\Big[  \bm{\mathcal{H}}_{1 \alpha 1
\beta} - \bm{\mathcal{H}}_{1 \alpha 2 \beta} + \bm{\mathcal{H}}_{2 \alpha 1
\beta} - \bm{\mathcal{H}}_{2\alpha 2 \beta}  \Big]  \cdot \m f_{\beta}.  
\label{eq:c-a}
\end{align}
Here 
 $\bm{\mathcal{H}}_{p \alpha q \beta} := \bm{H}(\m x^{\alpha}_p - \m x^{\beta}_q)$ 
is defined for $\alpha \neq \beta$  and $r \gg l,a$.
It is convenient to define 
\begin{align}
\m r  & := \m R_{A} - \m R_{B}\nonumber;\\
\m R  & := \frac{1}{2}(\m R_{A} + \m R_{B}) ;
\label{eq:trasf-2}
\end{align}
the relative and mean position respectively of the two centres of rotors A and B.
The magnitude  $r$ of
the relative distance $\m r$ 
is the typical separation between the objects.
For
 large separations, 
 $r \gg L$, it is natural to expand the hydrodynamic tensor using the small parameter $\epsilon : = L/r$.

\subsection{Angular dynamics}
From eqns.~(\ref{eq:c-a}),  it is clear that angular
and rotational dynamics are coupled. 
The equation for the director $\hh u_{\alpha}$
shows that the angular velocity receives a  correction coming from the
interaction with the other rotor. 
{It 
 is small (of order $\epsilon^2$)}
 compared to the 
leading order $O(1)$  term, which is unperturbed rotation of $\hh u$ with frequency 
$\omega_0$.
{It can be shown perturbatively to be a time-dependent effect with
zero average value over a cycle to any order of $\epsilon$. Thus, it does not
induce a shift in $\omega_0$~\cite{LL2011}.} 
Therefore,  
{
\begin{align}
& \hh u_{\alpha}  = \hh x \sin(\sigma_{\alpha } \omega_0 t ) + \hh y
\cos(\sigma_{\alpha } \omega_0 t )  + O(\epsilon^2) \; , \nonumber\\
& \nonumber\\
&\hh u_{\alpha}^{\perp} = \hh x \cos(\sigma_{\alpha } \omega_0 t )  - \hh y
\sin(\sigma_{\alpha } \omega_0 t )  + O(\epsilon^2) \; . 
\label{eq:u-unperturb}
\end{align}
}
Since the isolated rotors do not translate, the  first non-zero contribution to the translational motion is at least  of order $\epsilon$. Hence to leading order in $\epsilon$, the rotational motion affects translational motion of the rotors but {\em  not} vice-versa.
In other words, the individual  rotor directors $\hh u_{\alpha}$  are fast variables  and their positions $\m
R_{\alpha}$ are slow variables.  Consequently,  we are now in the position to study the effects of the fast degrees of freedom on  the (slow) dynamics.

\subsection{Translational dynamics}

The rotor directors $\hh u_\alpha$ are periodic functions of period $T_0$.
To study the dynamics of the position $\m R_\alpha$ on time-scales much greater than the period, $t \gg T_0$, we use a dynamic variant of asymptotic homogenization~\cite{pavliotis}. This involves performing an asymptotic expansion in powers of the ratio of the short time-scale to the long time-scale, $(t/T_0)^{-1}$, while implementing the condition that the microscopic dynamics is periodic.  
From this, given an initial configuration: $ \left\{{\bf R}_\alpha(0)\right\}$,  we obtain an `averaged'  equation~\cite{strogatz} for  the velocity over  a time period $T_0$, $\overline{\bf v}_\alpha = \frac1T_0\int_0^{T_0} dt \;  \partial_t \m R_{\alpha} $ for the rotor centres, which we denote by an over-bar.
This is equivalent to calculating  the displacement in a period $T_0$,   $\Delta \m R_{\alpha}(T_0)$ obtained by integrating eq.~(\ref{eq:c-a}) over a period $T_0$, where orientations are given by eq.~(\ref{eq:u-unperturb}) and dividing the displacement by the period $T_0$.

\begin{figure}[!ht]
\centering
\includegraphics[width=0.45\textwidth]{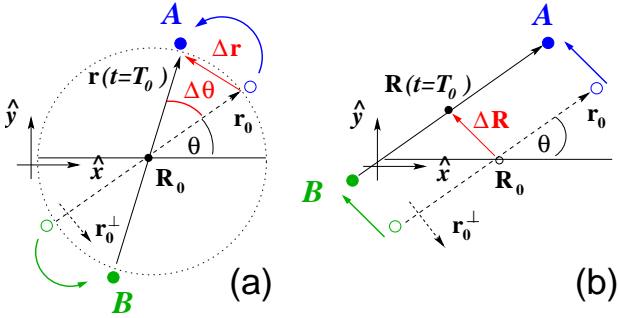}
\caption{ 
{Slow dynamics of the rotor centres.}
 (a) Equal rotors. The mean position $\m R$ is at rest and the dynamics
involves the relative distance $\m r$ only. 
After a  period $T_0$, we  observe a net rotation of an angle $\Delta \theta$. Here $\sigma_A = \sigma_B = -1$
(b) Opposite rotors. In this case $\m R$ moves and $\m r$ 
{on average stays at its initial value.
}
 After a period $T_0$, there is a net translation of $\m R$ along 
the direction $\hh r^{\perp}_0$. Here $\sigma_A =1$.} 
 \label{fig:strobo-disp-analytic}
\end{figure}

As above, it is useful to use the relative position, $\m r$ and mean position, $\m R$, as defined in eqn. (\ref{eq:trasf-2}).
We use a subscript 0  for the positions at the instant $t = 0$, so  
$\m R_0 \equiv \m R(t = 0)$ and  
$\m r_0 \equiv \m r(t = 0)$. 
As one would expect the translational dynamics depends on the relative spin of the two rotors.
To illustrate this we define $\sigma_{\pm} := 1/2(\sigma_A \pm \sigma_B) $:  $\sigma_+$ is non-zero for equal rotors and zero for opposite rotors while $\sigma_-$ is zero for equal rotors and non-zero for opposite rotors. 
Then
{   to leading order in $\epsilon$ }
 we find:
\begin{enumerate}[(i)]
\item for equal rotors ($\sigma_A = \sigma_B$) 
$\mathbf{R}$ is left fixed and $\m r$ rotates around it with a mean angular
velocity 
\begin{equation}
\overline{\bm \omega_1} := \dfrac{1}{T_0} \int^{T_0}_0  -\dot{\theta} \hh z dt =  
\omega_1 \sigma_+ \hh z 
\label{eq:omega1}
\end{equation}
as shown in Fig.~\ref{fig:strobo-disp-analytic} (a).  $\theta$ is the angle
formed by $\m r$ with the $\hh x$ axis that increases in the counter-clockwise sense and $ \omega_1 = ( l f \epsilon^3_0)/(2 \pi \eta L^2) = (\omega_0 l \epsilon_0^3)/(L g(l/a)) $,  introducing the time-scale $T_1 := 2 \pi/ \omega_1$. Here $\epsilon_0 = L/r_0$ where $r_0$ indicates the magnitude of $\m r_0$. Due to the dilute nature of the solution we have $T_0 \ll T_1$, hence  $\m r$  rotates much slower than  $\hh u_{\alpha}$. 
{This}
  is in qualitative agreement with previous work~\cite{LJJP,Pago} using different rotor models;

\item for opposite rotors ($\sigma_A = -\sigma_B$) $\m r $ 
{ remains constant on average } 
and $\m R$ moves  orthogonal to $\m r$  with a mean velocity
\begin{equation}
\overline{ \dot{\m R}} := \dfrac{1}{T_0} \int^{T_0}_0  \dot{\m R} dt= - v_1
\sigma_- \hh r^{\perp}_0 
\label{eq:v1}
\end{equation}
as shown in Figure~\ref{fig:strobo-disp-analytic} (b). Here $ v_1 = r_0 \omega_1/2$ and  we identify $\hh r^{\perp} = \hh r^{\perp}_0$, since
$\m r$ is on average constant. In contrast to identical rotors at low Re~\cite{LJJP,Pago}, this problem has not received much attention. 
\end{enumerate}

We consider the intrinsic reference frame described by unit vectors
$\hh r_0 $, $\hh r^{\perp}_0$ (which are defined in terms of the vectors $\m R_A, \m R_B$ at $t=0$).
They satisfy $ \hh r_0  \wedge \hh r^{\perp}_0 = \hh z$.  %
The results above can be expressed in terms of the velocity of the centres  of
each rotor.  The {\em slow} dynamics of rotor A, can be expressed  compactly by  the velocity
{ 
\begin{equation}
 \dot{\mathbf{R}}_{A} (t) =  \dfrac{1}{2}\Big( \overline{\bm \omega_1} \wedge \m
r(t)  \Big) +  \overline{ \dot{\m R}}  , 
\label{eq:average-vel}
\end{equation}
}
with the corresponding expression for rotor B. 
This represents either the effect of a net rotation, occurring only when the rotors have equal spins, or a net translation, if they have opposite spins.
Since for a pair of equal rotors, $\m r$ rotates, it is convenient to use the parametrization
$ \m r(t) = r_0 \Big( \cos  \Omega(t)  \hh r_0 + \sin \Omega(t)  \hh r^{\perp}_0
 \Big)$
where  $\Omega(t) :=  \omega_1 \sigma_+ t$.

 {
 It is interesting to note that symmetry considerations alone indicate that the qualitative nature of the motion of a pair should be independent of the detailed form of rotor~\cite{Pago,LJJP}.
If the separation of the rotors is large compared to their size, the net flow is well approximated by 
superposing the flow fields generated by the individual rotors.  
On timescales long compared to the period of the rotation of $\hh u$, their combined motion can only depend on the two directions $\hh r_0$ and $\hh r^\perp_0$ characterising the configuration of the rotors.
Moreover, the flow field  generated by each rotor has rotational symmetry.
From this it follows that  any net  velocity  has no component on the  $\hh r_0$ direction. 
Hence, it can only have a net component  along the $\hh r^\perp_0$ direction. Implementing this for  two \emph{identical} rotors, one easily sees that  their two  centres are convected by equal and opposite flows - leading to rotation;  while when they have \emph{opposite} spin they are both subject to the same velocity and translate together. This insight can be very productively combined with analytic techniques to investigate the slow dynamics.
 }

{ \subsection{  Comparison with numerics} 

\begin{figure}[h!]
\centering
\includegraphics[width=0.48\textwidth]{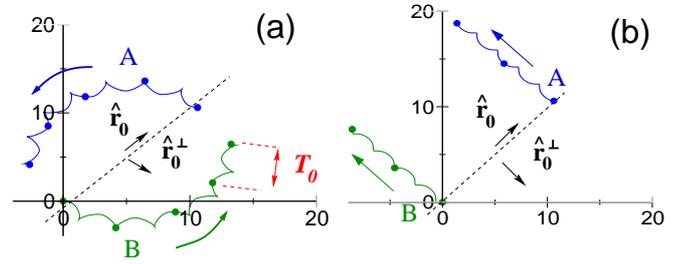}
\caption{ {Dynamic trajectories of rotor centres in the plane, obtained from numerical simulations.
Continuous lines represent the instantaneous positions of the rotor centres incorporating \emph{both}  `slow' and `fast' motions; solid circles show its slow envelope,  at  time intervals that are multiples of the  period $T_0$.
 (a) Equal rotors with $\sigma_A =-1$.   
 The trajectories of the  rotors oscillate in the radial direction while performing a revolution around their common centre. 
  The slow dynamics, however,  describes a circular trajectory. (b) Opposite rotors, with $\sigma_A =1$. The two rotor centres  oscillate around the radial direction while  moving on average in a direction orthogonal to  their separation: 
   the `slow' motion is a straight line in this direction. 
The initial relative orientation is $\hh r_0 = \hh x + \hh y$ and the parameters  are $l= 5 \mu$m, $a=0.5 \mu$m, $f=10$pN, $L =6 \mu$m,  $r_0 =15 \mu$m. }}  
\label{fig:traject-2rot-numeric}
\end{figure}

We compare analytical results with numerical simulations that involve the full tensorial expression of the hydrodynamic tensor, allowing us to investigate the dynamics of two rotors beyond the regime where $\epsilon \ll 1$. 
Equations~(\ref{eq:c-a}) are implemented  in C and integrated 
numerically using  the Taylor algorithm~\cite{CLJ03}.  
The accuracy of the results can be improved by simply using a smaller integration step $dt$.  
  Here, we have used a time-step $dt =10^{-4} \frac{T_0}{4 }$. We consider many internal cycles $T_0$  and keep track of positions and angles of rotors both  i)  at each time step  (instantaneous)   and ii)  after a period $T_0$. 
We have checked that a different phase prescription for $\hh u^\alpha$ does not affect the dynamics, 
which is qualitatively unchanged even in the case of a phase difference between the directors.

\begin{figure}[h!]
\centering
\includegraphics[width=0.48\textwidth]{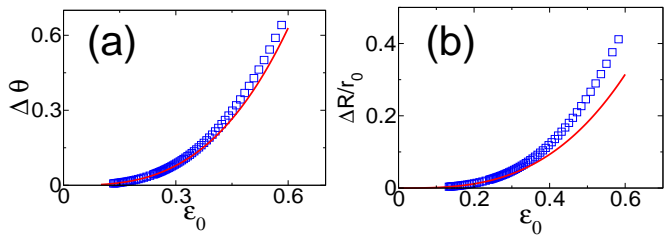}
\caption{  { Net changes in a cycle as functions of $\epsilon_0$.  Approximated analytical results (continuous curves) compared with numerical simulations (squares).  (a)  Equal rotors. The net angle $|\Delta \theta|$ in a cycle.  (b) Opposite rotors. Net displacements $|\Delta R|$ in a cycle, rescaled with the initial separation $r_0$. In both cases parameters are the same of figure~(\ref{fig:traject-2rot-numeric})  (except  $r_0$).  }}  
\label{fig:comparison-numeric}
\end{figure}
 
For $\epsilon \ll 1$, results are in quantitative agreement  with approximate analytical solutions,
with which they are compared in
Figure~(\ref{fig:comparison-numeric}) as a function of the initial separation $r_0$, at constant $L$. 
As $\epsilon$ is increased,  deviations appear as expected, however the qualitative behaviour remains unchanged.
There we plot the net rotation angle after a period  $T_0$ from simulations and the approximate analytic result given by
\begin{equation}
 | \Delta \theta | = T_0 \omega_1 = \frac{2 \pi l \epsilon^3_0  }{L \, g\left(l/a\right) } 
 \label{eq:delta-theta}
\end{equation}
 and similarly  the net displacement, rescaled with  $r_0$, compared with the approximate analytic result
\begin{equation}
\left| \frac{\Delta R}{r_0} \right|= T_0 v_1 = \frac{ \pi l \epsilon^3_0 }{L \, g\left(l/a\right) }. 
\label{eq:delta-c}
\end{equation}
 
\subsection{Comparison to three dimensional hydrodynamics}
It is also instructive to consider (briefly) rotors that move in an unbounded three dimensional fluid of viscosity $\eta$.  Here rotors are composed of spheres of radius $a$, rather than disks, linked together. We assume that their directors $\hh u^\alpha$  can be confined  to two dimensions by some  mechanism. The dynamics of point forces $\m f = f \sigma \hh u^\perp$ acting on the spheres and lying entirely in that plane can be represented by means of the Oseen tensor~\cite{Doi}: $\bm{H}(\m r) =\frac{1}{8 \pi \eta} \frac{\mathbb{I} + \hh r \otimes \hh r
}{r}$ for $r\gg a$. The  phenomenology that we find is qualitatively similar to rotors moving in a thin film, with only qualitative differences.
The centre of a single rotor is at rest and its director $\hh u$ rotates with a constant angular velocity $\omega^{(3d)}_0 = \frac{f}{\pi \eta L }(\frac{1}{3a}-\frac{1}{4L})$. 
Two interacting rotors show again a combination of fast and slow dynamics. The slow motion is a rotation around the common centre for rotors having the same spin, and a net translation for rotors having opposite spin. These slow motions are  described again by means of equations that have the same form of eqs~(\ref{eq:omega1}) and~(\ref{eq:v1}), with same directions but  different magnitudes.  To leading order in $\epsilon$ these magnitudes are $\omega^{(3d)}_1 = \frac{ \epsilon^3_0 f}{4 \pi \eta L^2}$,  for same rotors
 and $v^{(3d)}_1 = \frac{1}{2} r_0 \omega^{(3d)}_1$, for opposite rotors.

Similar results can also be obtained also by replacing the model of rotor with a single spinning sphere of radius $b$ with an angular velocity  $\bm{\Omega}$ of fixed magnitude, directed along the $\pm \hh z$ axis.  The flow field  at distance $r$, generated by a sphere in placed at the origin, is  given by~\cite{Lifshitz:1987jw}
\begin{equation}
\m v = \frac{b^3}{r^3}\bm{\Omega} \wedge \m r.
\label{eq:spin-sphere}
\end{equation}
For separations large compared to the sphere radius, the velocities of the  two sphere centres can be approximated by superposing the fields generated by each sphere. Hence for rotors of like spin,  i.e.  $\bm{\Omega}_A = \bm{\Omega}_B$,  the motion of the pair is a rotation about their common centre with angular velocity given by $\frac{1}{2} \m r \wedge \m v$, with $\m v$ the same as in eq~(\ref{eq:spin-sphere}). In contrast, for opposite spin, each sphere translates in the direction perpendicular to the separation vector $\m r$, with a  velocity given by~(\ref{eq:spin-sphere}). 
} 
{
\subsection{A pair of rotors as a model swimmer}
 The dynamical behaviour of two rotors  described above suggests a natural application.  
 Since two opposite rotors move in a straight line,
  they provide a nice model of  a  self-propeller, whose force density has the form of a \emph{torque dipole}. 
   The  mean propulsion  speed depends on the rotor  separation and follows  from 
   eqn.~(\ref{eq:delta-c}) as $\hat v = \Delta R / T_0$.
   The direction of motion  can be varied easily by controlling the torques: the motion can be reversed  by  simultaneously inverting both the torques.  The swimmer can be made to change its direction by  reversing  one of them for a short time since as
   discussed above, when the two spins have the same value $\sigma = 1$ ( $\sigma = -1$)  the swimmer will rotate clockwise (counter-clockwise) with an 
   angular displacement (in 1 cycle) given to leading order by eqn.~(\ref{eq:delta-theta}).

Taking account of fluctuations will lead to Brownian motion along the direction of the rotor separation with a diffusion constant $D$ proportional to the magnitude of the fluctuations.
This means that the speed of the propulsion which will fluctuate around the average $\hat v$. As a result, our description will remain valid until the separation of the rotors becomes comparable to rotor size  $r_0 \sim  L$, when the rotors strongly interact with each other. 
This  can be used to estimate the lifetime of the rotor-swimmer as a first passage time  off a particle diffusing in one dimension~\cite{Redner}.
To avoid this, and obtain a permanent swimmer,  two rotors could be mechanically linked together a fixed distance apart by a rigid linker~\cite{livmag01}.
 
Finally we make some general observations about  the properties of this self-propeller.
As required by the scallop theorem~\cite{Purcell}, a low Reynolds number swimmer 
must undergo cyclic shape deformations that are non-reciprocal requiring 
at least two degrees of freedom. It is clear that the rotations of  the pairs of rotors prescribe a non-reciprocal cyclic motion.
So despite being force-free, and torque-free, an assembly made of two rotors can swim, as a result of hydrodynamic interactions. 
It provides another example of an idealised simple swimmer model~\cite{NG04,bartolo,NZ10}  that it might be possible to construct on the nanoscale. 
A possible experimental realisation would be a pump created by dielectric spheres made to rotate by  a multi-bead optical trap.
It is noteworthy that more complex rotor-swimmers composed of two torque dipoles~\cite{LJJP} of opposite sign can also be used to 
construct an albeit less efficient swimmer~\cite{LL2011}.
}

In conclusion, we have investigated analytically and numerically the 
dynamics of rotors confined to two
dimensions by a viscous film (such as a membrane) sandwiched between a less viscous bulk fluid.
We first introduced the model of rotor and subsequently 
studied the  interaction between two such rotors.
We 
 found that the resulting dynamics
depends on the relative spin of the pair
   and that it involves fast and slow quantities.  Concentrating on the slow variables, we find that 
 for opposite rotors the hydrodynamic interactions lead to a net translation of the pair while for like rotors they lead to a relative rotation of the pair around each other.  
There are  a number of possible experimental realisations of this system. For example,  qualitative studies of  chemically driven 'nano'-rotors have already been performed. A promising direction for future studies is to confine them in thin films giving greater control of their behaviour and allowing a more quantitative study of their interactions.
%
{  
Another natural application is self-locomotion. Two rotors can be used to construct a swimmer: when the torques are equal and opposite, the swimmer  translates;  when the torques are the same, it can rotate in a controlled way with respect to its original direction, as discussed above. 
Alternatively, two tethered rotors with opposite spin can generate a net flow along their common axis, while rotors with same spin would generate a flow with rotational symmetry around them. 
Finally, this study
serves as  basis 
and   provides a natural framework 
  to describe the  collective behaviour of many rotors, 
that we will  address in the future~\cite{LL2011}.
  }

\acknowledgments

ML acknowledges the support of University of Bristol research studentship.
TBL acknowledges the support of the EPSRC under under grant EP/G026440/1.
{  ML wishes to acknowledge useful conversations with Eva Baresel.}
We thank M.C. Marchetti for illuminating discussions.


\end{document}